\documentclass[11pt]{article}
\usepackage{amsmath,amsfonts}
\allowdisplaybreaks
\def \beq  {\begin{equation}}
\def \eeq  {\end{equation}}
\def \beqar {\begin{eqnarray}}
\def \eeqar {\end{eqnarray}}

\def\be{\begin{equation}}
\def\ee{\end{equation}}
\def\bsp{\be\begin{split}}
\def\sqr#1#2{{\vcenter{\vbox{\hrule height.#2pt
\hbox{\vrule width.#2pt height#1pt \kern#1pt
\vrule width.#2pt}\hrule height.#2pt}}}}

\def\del {\partial}

\begin{document}
\fontfamily{cmr}\fontsize{12pt}{15pt}\selectfont
\def \CMP {{Commun. Math. Phys.}}
\def \PRL {{Phys. Rev. Lett.}}
\def \PL {{Phys. Lett.}}
\def \NPBProc {{Nucl. Phys. B (Proc. Suppl.)}}
\def \NP {{Nucl. Phys.}}
\def \RMP {{Rev. Mod. Phys.}}
\def \JGP {{J. Geom. Phys.}}
\def \CQG {{Class. Quant. Grav.}}
\def \MPL {{Mod. Phys. Lett.}}
\def \IJMP {{ Int. J. Mod. Phys.}}
\def \JHEP {{JHEP}}
\def \PR {{Phys. Rev.}}
\def \JMP {{J. Math. Phys.}}
\def \GRG{{Gen. Rel. Grav.}}
\begin{titlepage}
\null\vspace{-62pt} \pagestyle{empty}
\begin{center}
\vspace{1truein} 
\vskip .1in
{\Large\bfseries
Holomorphicity, Vortex Attachment, Gauge Invariance and the Fractional Quantum Hall Effect}\\
\vspace{6pt}
\vskip .1in
{\Large \bfseries  ~}\\
\vskip .1in
{\Large\bfseries ~}\\
{\large\sc Abhishek Agarwal$^{a,b}$}\\
\vskip .2in
{\itshape $^a$\,Physical Review Letters\\
American Physical Society\\\
Ridge, NY 11961}\\
\vskip .1in{\itshape $^b$\,Department of Physics and Astronomy\\
Lehman College of the CUNY\\ 
Bronx, NY 10468}\\
\vskip .1in
\begin{tabular}{r l}
E-mail:
&{\fontfamily{cmtt}\fontsize{11pt}{15pt}\selectfont abhishek@aps.org}
\end{tabular}

\fontfamily{cmr}\fontsize{11pt}{15pt}\selectfont
\vspace{.8in}
\vspace{1.5in}

\centerline{\large\bf Abstract}
\end{center}
A gauge invariant reformulation of nonrelativistic fermions in background magnetic fields is used to obtain the Laughlin and Jain wave functions as $exact$ results in Mean Field Theory (MFT). The gauge invariant framework trades the $U(1)$ gauge symmetry for an emergent holomorphic symmetry and fluxes for vortices. The novel holomorphic invariance is used to develop an analytical method for attaching vortices to particles. Vortex attachment methods introduced in this paper are subsequently employed to construct the Read operator within a second quantized framework and obtain the Laughlin and Jain wave functions as exact results entirely within a mean-field approximation. The gauge invariant framework and vortex attachment techniques are generalized to the case of spherical geometry and spherical counterparts of Laughlin and Jain wave functions are also obtained exactly within MFT.
\end{titlepage}
\pagestyle{plain} \setcounter{page}{2}
\fontfamily{cmr}\fontsize{12pt}{15.5pt}\selectfont
\section{Introduction and Summary}
The complex analytic structure of the iconic Laughlin wave function\cite{RL}
\be
\mathcal{U}_e = \prod_{i< j} (\bar z_i - \bar z_j)^{2s+1}e^{-\sum_i\frac{B}{4}z_i\bar z_i} \label{laughlin}
\ee
strongly suggests that its natural habitat is a quantum mechanical framework where holomorphicity plays a key role\footnote{In our notation $\bar z = x+iy$, $z = x-iy$.}. The same expectation also applies to generalizations of the Laughlin wave function to $\nu = \frac{p}{2ps+1}$: namely the Jain states\cite{JJ}.
\be
\mathcal{U}_e^p =    e^{-\frac{B}{4}\sum_i z_i\bar z_i } P_{LLL}\left[\prod _{i<j}(\bar z_i - \bar z_j)^{2s}\tilde{\mathcal{U}}_{\Phi}^{p, B^*} (\{\bar z_i\}, \{z_i\})\right] \label{jain}
\ee
In the above expression $\tilde{\mathcal{U}}_{\Phi}^{p, B^*}$ denotes the Landau Level (LL) wave function (without the gaussian factor) for $p$ filled composite fermion LLs in a magnetic field $B^* = \frac{B}{2ps+1}$, while  $ P_{LLL}$ is the projection operator to the Lowest Landau Level (LLL)\footnote{As will be clear in later discussions, the label $\Phi$ refers to field operators for composite particles.}. It is clear that  holomorphic properties of the wave functions underlie much of the interesting physics - including localization and compressibility properties -  that follow from them. Further, as is well known, these wave functions can be related to the electronic vacuum state via a nonlocal transformation\cite{read}. The operator that accomplishes this map - the Read operator - is complex and not unitary. Symmetries following from the Read operator - $W_\infty$ algebras -  are also manifestly complex without a real counterpart\cite{winfty}. All these attributes of Laughlin and Jain wave functions further bolster the intuitive notion that a holomorphic framework must underlie the dynamical set up for the fractional quantum Hall effect (FQHE). \\
However, the standard field theoretic routes to  obtaining (\ref{laughlin},\ref{jain}) are based on flux attachment techniques which - in turn - utilize the unitary gauge symmetry of electronic dynamics. The gauge fields, and more importantly, the parameter of the underlying $U(1)$ internal gauge symmetry are both real. As has long been appreciated, attempts at deriving (\ref{laughlin},\ref{jain}) based on unitary symmetries are bound to encounter various incongruities. For example, the physical picture presented by (\ref{laughlin},\ref{jain}) is one of particle degrees of freedom bound to vortices, not fluxes. Vortices - described by the Jastrow factor ($\prod_{i<j} (\bar z_i - \bar z_j)^{k}$)  - are naturally complex valued and their attachment to particles  is hard to incorporate in any direct fashion within  dynamical frameworks based on unitary symmetries.
To better appreciate the tension between real symmetries and complex vortices it is helpful to recall some technical details. The flux attachment procedure maps the electronic wave function $\mathcal{U}_e$ to that of a composite boson or fermion $\mathcal{U}_\Phi$ as follows.
\be
\mathcal{U}_e = \prod_{i<j}\left[\frac{\bar z_i - \bar z_j}{z_i - z_j}\right]^{m/2}\mathcal{U}_\Phi\label{fluxw}
\ee
This is a unitary transformation  - albeit a singular one - and it is consistent with the unitary gauge symmetry of electronic dynamics. (\ref{fluxw}) attaches m units of flux to each particle. The fluxes translate to a statistical gauge field in the hamiltonian which can be  diagonalized exactly at the level of mean field theory (MFT)\footnote{We review flux attachment techniques in a second quantized formalism in some detail in the next section.}. The resultant mean field wave functions are:
\be
\mathcal{U}_e (\{z_i\}, \{\bar z_i\}) = \prod _{i< j}\left[\frac{\bar z_i - \bar z_j}{z_i - z_j}\right]^{(2s+1)/2}\label{mft} 
\ee
when $p=1$ and and $m = 2s+1$ (the composite particles are bosonic). And
\be
\mathcal{U}_e (\{z_i\}, \{\bar z_i\}) = \prod _{i<j}\left[\frac{\bar z_i - \bar z_j}{z_i - z_j}\right]^{s} \mathcal{U}_{\Phi}^{p,B^*}(\{z_i\}, \{\bar z_i\})\label{mftp}
\ee
when $p \geq 1$ and $m=2s$ (the composite particles are fermionic). In our notation, $\mathcal{U}_{\Phi}^{p,B^*}(\{z_i\}, \{\bar z_i\})$ represents the LL wave function  - including the gaussian factor - with $p$ fully occupied levels at background magnetic field $B^*$, while $\tilde{\mathcal{U}}_{\Phi}^{p,B^*}(\{z_i\}, \{\bar z_i\})$ represents the same without the gaussian factor ($\mathcal{U}_{\Phi}^{p,B^*} = e^{-\sum_i\frac{B^*}{4}z_i\bar z_i} \tilde{\mathcal{U}}_{\Phi}^{p,B^*}$)\footnote{e.g. when $p=1$, $\mathcal{U}_{\Phi}^{p,B^*}(\{z_i\}, \{\bar z_i\})$ = $\prod_{i< j} (\bar z_i - \bar z_j)^{2s+1}e^{-\sum_i\frac{B^*}{4}z_i\bar z_i} $ and $\tilde{\mathcal{U}}_{\Phi}^{p,B^*}(\{z_i\}, \{\bar z_i\})$ = $\prod_{i< j} (\bar z_i - \bar z_j)^{2s+1}$}.
Clearly (\ref{mft}) and (\ref{mftp}) are a far cry from the the desired wave functions (\ref{laughlin}) and (\ref{jain}). Apart from statistics, the mean field wave functions share little in common with (\ref{laughlin}) and (\ref{jain}).  Crucially, the zeroes and the correct gaussian factors of  (\ref{laughlin}) and (\ref{jain})  are both absent in (\ref{mft}) and (\ref{mftp}). This shortcoming can fundamentally be traced back to the fact that the``parameters" in (\ref{fluxw}) could only be chosen to capture the unitary part of the zeroes of the wave functions  - namely the phases - to preserve overall gauge invariance of the underlying dynamics.  What one needs instead is the attachment of vortices ( $\prod_{i<j} (\bar z_i - \bar z_j)^{2s+1}$) which cannot be done directly within the usual gauge covariant set up (reviewed in the next section) without jeopardizing gauge invariance. One is tempted to guess that vortex attachments would be natural in a reformulation of the original theory where holomorphic transformations, instead of $U(1)$ rotations, played the role of internal symmetries.\\
The standard procedure for improving upon (\ref{mft},\ref{mftp}) involves going beyond MFT and including the effect of fluctuations \cite{zhang,fradkin,kane,shankar}. Since we shall work within a hamiltonian framework, it is worth drawing special attention to the remarkable series of papers by Shankar and Murthy \cite{shankar} where a Bohm-Pines approach was used to incorporate effects of fluctuations to recover (\ref{laughlin}) and (\ref{jain}) by computing corrections to the MF wave functions (\ref{mft},\ref{mftp}) \footnote{See \cite{shankar-book} for a detailed pedagogical review of this work. Some essential features of the Bohm-Pines approach - adapted to the notation of this paper - are also recapitulated in the Appendix.}. Despite the clear success of this approach in obtaining FQH wave functions on the plane, several open issues remain. For example, it is not clear if the departure from MFT is characterized by a natural expansion parameter. The absence of a parameter to control the size of fluctuations limits the amount of available analytical control. More importantly, we are still left without a clear mathematical framework for attaching vortices (as opposed to fluxes) to particles. Further it is not immediately clear how readily the path integral\cite{fradkin} or hamiltonian\cite{shankar} approaches to incorporating fluctuations generalize to nontrivial geometries or to nonabelian theories.\\ 
Is there an alternative route to (\ref{laughlin}, \ref{jain}) that may help us address the issues left open by flux attachment techniques? A careful examination of the analysis done in \cite{fradkin, shankar} shows that at no point were any details specific to the system (e.g. material properties, band structure etc) used in the derivation of the planar FQH wave functions. One is then led to surmise that the role of fluctuations in \cite{fradkin, shankar} is simply to resum generic physical properties of the system which were missed by MFT due to the choice of the original degrees of freedom and the concomitant unitary gauge symmetry. This suggests that a better reorganization of the physical degrees of freedom may help us recover the FQH wave functions exactly via MFT,  rendering the resummation accomplished by density fluctuations unnecessary.\\
Such a reorganization of the degrees of freedom is indeed  part of what we present in this paper. We begin by reformulating the dynamics of nonrelativistic matter coupled to Maxwell fields in a manifestly gauge invariant form. We do so using the framework developed by Karabali, Kim and Nair (KKN) which is well suited for  hamiltonian analyses of gauge theories in D=2+1\footnote{The KKN formalism was first used to understand the nonperturbative mass-gap\cite{KKN1} and string tension\cite{KKN1,KKN2} of purely gluonic nonabelian Yang-Mills theory. Subsequently, it has  been employed in nonperturbative studies of QCD with and without supersymmetry in three spacetime dimensions\cite{AKN}. For more recent applications of this framework to studies of entanglement entropy and fermion-fermion dualities in D=2+1, see\cite{AKN-EE, AAL}.}. The gauge invariant reformulation  trades the $U(1)$ gauge symmetry for an emergent holomorphic symmetry. Denoting the gauge invariant version of the second quantized electronic field by $\Lambda$, the holomorphic symmetry acts as
\be
\Lambda \rightarrow \bar{V}(\bar z) \Lambda
\ee
where $\bar{V}$ is an arbitrary function that depends only on the holomorphic coordinate $\bar z$ (and not on $z$). This new symmetry allows us to trade flux attachments (which are natural for a theory where the internal symmetry is parameterized by phases) for vortex attachments (which we show to be natural in the reformulated theory with an internal holomorphic symmetry). This trade off has several important consequences. At a conceptual level, the gauge invariant formalism and the associated holomorphic symmetry provides us a with a natural mathematical framework for attaching vortices to matter fields. In other words we are able to construct a second quantized form of the Read operator. At a computational level, our approach allows us to recover both (\ref{laughlin}) and (\ref{jain}) as $exact$ results in mean field theory. The construction of the Read operator and the derivation of the wave functions utilizes only (a gauge invariant combination of) the physical electronic and photonic degrees of freedom. We are thus able to recover (\ref{laughlin}) and (\ref{jain}) without needing either to enlarge the hilbert space (as was the case in \cite{shankar} ) or go beyond MFT. Further, the wave functions derived within the gauge invariant framework are naturally in the Bargmann \cite{girvin} form from which the $P_{LLL}$ operation follows naturally. \\
As further illustration of the utility and robustness of our framework we generalize our study to spherical geometry, where regarding $S^2$ as the coset space $SU(2)/U(1) \equiv CP^1$ simplifies matters considerably. The coset space structure of $CP^k$ spaces has been extensively used to study the integer quantum Hall effect on $CP^1$\cite{KN1} and higher dimensional projective spaces\cite{KN1,KN2}. For the specific case of $CP^1$, a gauge invariant hamiltonian reformulation of nonabelaian gauge theories is also known \cite{AN2}. Combining the frameworks of \cite{AN2} and \cite{KN1} gives us the tools necessary to formulate vortex attachments on $S^2$ using an emergent holomorphic symmetry. As a result we are able to derive the spherical counterparts of (\ref{laughlin}) and (\ref{jain}), once again, as exact results within MFT. \\ 
It bears mentioning that Rajaraman and Sondhi - in an insightful paper \cite{RS} - proposed complexifying $U(1)$ gauge transformations (\ref{fluxw}, \ref{fluxa}) and the associated statistical gauge field. Under a specific choice of gauge fixing, this construction produced the Laughlin wave function on the plane exactly at MFT. However, since the complexified gauge transformation is not obtained from an underlying internal symmetry,  it is not immediately clear that the resultant composite particle hamiltonian is gauge invariant in general. Further, generalizability of this approach to nonplanar geometries or to nonablelian theories is an open issue. Nevertheless, the highly suggestive results in \cite{RS} underscore the expectation that a complex generalization of $U(1)$ symmetries may provide a more economical and direct route to FQH wave functions.
The present paper provides a symmetry based rationale for complexified (holomorphic) gauge transformations. The complex valued nature of the Read operator emerges naturally in our approach as a consequence of the holomorphic symmetry underlying the gauge invariant framework. Having an underlying symmetry as a guiding principle makes the present approach generalizable to non-planar geometries. 
Further, the new holomorphic symmetry - as we show - is exactly what is needed for vortex attachments. \\
This paper is organized as follows. We start with a brief review of flux attachment methods adapted to the second quantized hamiltonian set up of this paper. In particular, we derive ({\ref{mft} and \ref{mftp}) and illustrate how flux attachment fails to capture FQH physics at the level of MFT. Following this discussion we introduce the gauge invariant hamiltonian framework and elaborate on how an emergent holomorphic symmetry follows from a gauge invariant reorganization of electronic dynamics. In the following sections we apply the gauge invariant formalism to FQH effects on $R^2$ and $S^2$ respectively. In particular, we utilize holomorphic invariance to obtain a field theoretic method for attaching vortices to gauge invariant dynamical degrees of freedom. We then formulate a holomorphic MFT appropriate to  the gauge invariant framework and derive both planar and spherical Laughlin and Jain wave functions as exact result within a holomorphic MFT. We end the paper with some concluding thoughts and prospects for future developments. The Appendix is devoted to a review of the Bohm-Pines methods developed in \cite{shankar}. In particular, we argue that the Bohm-Pines method applied to the results of the gauge invariant formalism does not produce further corrections to the holomorphic MFT results.  
\section{Flux Attachment and Second Quantization: A Brief Review}
We begin with an abridged review of flux attachment techniques in a second quantized framework to better expose the tension between unitary gauge symmetries and the inherently complex nature of FQHE states. 
In the absence of interactions, the  second quantized electronic hamiltonian in a constant background magnetic field is\footnote{$\del = \frac{1}{2}(\del_1 + i \del_2)$, $\bar \del = \frac{1}{2}(\del _1  - i\del_2)$. }
\be
H_e = -\frac{2}{\mu}\int \Psi_e^\dagger(\bar \partial - \frac{z}{4}B)(\partial + \frac{\bar z}{4}B)\Psi_e \label{He}.
\ee
The non-relativistic fermion field satisfies
\be
\{\Psi_e (\vec{x}), \Psi^\dagger_e(\vec{x}')\} = \delta ^2(\vec{x}-\vec{x}'). 
\ee
Electrons can be transformed to composite particles $\Phi$ via a field dependent singular gauge transformation.
\be
\Psi_e(z,\bar z) = \left(\exp{\frac{m}{2}\int _w \ln \left[\frac{\bar z - \bar w}{z-w}\right]\rho(w,\bar w)}\right)\Phi(z, \bar z) \label{fluxa}
\ee
The parameter $m$ is an integer and the composite field $\Phi$ represents bosonic or fermionic particles for odd or even values of $m$ respectively\footnote{$\int_w$ is shorthand for $\int d(Re(w))d(Im(w))$}. $\rho = \Psi_e^\dagger \Psi_e$ is the density. The angular factor - $ \ln \left[\frac{\bar z - \bar w}{z-w}\right]$ - renders the above transformation unitary, as required by the unitary gauge invariance of the hamiltonian. However, since it is singular, (\ref{fluxa}) does change the physics by attaching magnetic fluxes to composite particles. The transformed hamiltonian
\be
H_e \rightarrow H_c = -\frac{2}{\mu}\int \Phi^\dagger(\bar \partial - \frac{z}{4}B + \bar a)(\partial + \frac{\bar z}{4}B + a)\Phi \label{Hp}
\ee
couples the composite particles to an additional statistical gauge field $a, \bar a$:
\be
\bar a = \frac{m}{2}\int _w \frac{\rho (w,\bar w)}{\bar z - \bar w}, \hspace{.2cm}   a = -\frac{m}{2}\int _w \frac{\rho (w,\bar w)}{z - w}
\ee
The constitutive relations 
\be
\partial \bar a  = \frac{m \pi}{2} \rho, \hspace{.2cm} \bar \partial a = -\frac{m \pi}{2}\rho \label{fluxpin} 
\ee
show that (\ref{fluxa}) has attached magnetic fluxes to the composite particles. One can set $\rho$ equal to its mean field value (the average number density), $\bar \rho$, in $H_c$ to derive a mean field theory (MFT) for composite particles. In particular, two specific relations between the external magnetic field $B$ and $\bar \rho$ are of special interest.\\
\be
m\bar \rho = \frac{B}{2\pi}, \hspace{.2cm} m = 2s+1, \hspace{.2cm} s = 1,2,3\cdots \label{cblevel}
\ee 
\be
\frac{\bar \rho}{\nu} =   \frac{B}{2\pi}, \hspace{.2cm} \nu = \frac{p}{2ps+1}, \hspace{.2cm} m=2s, \hspace{.2cm} p,s = 1,2,3\cdots \label{cflevel} 
\ee
The two cases render the composite particles bosonic and fermionic respectively. The mean field hamiltonian in both cases is.
\be
H_c = -\frac{2}{\mu}\int \Phi^\dagger(\bar \partial - \frac{z}{4}B^*)(\partial + \frac{\bar z}{4}B^*)\Phi \label{compH}
\ee
where the average magnetic field, $B^*$, experienced by the composite particles is:
\be
B^* = 0
\ee
for the composite boson case (\ref{cblevel}), while 
\be
B^* = \frac{B}{2ps+1}
\ee
in the case of the composite fermions (\ref{cflevel}). A general $N$ composite particle state 
\be
|N\rangle = \frac{1}{\sqrt{N!}}\int_{z_1...z_N}\mathcal{U}_\Phi (\{z_i\}, \{\bar z_i\})\Phi^\dagger(z_1,\bar z_1)...\Phi^\dagger(z_N, \bar z_N)|0\rangle
\ee
may equivalently be expressed as an electronic state via (\ref{fluxa})
\be
|N\rangle  = \frac{1}{\sqrt{N!}}\int_{z_1...z_N}\mathcal{U}_\Phi (\{z_i\}, \{\bar z_i\})\prod _{i< j}\left[\frac{\bar z_i - \bar z_j}{z_i - z_j}\right]^{m/2}\Psi_e^\dagger(z_1, \bar z_1)...\Psi_e^\dagger(z_N, \bar z_N)|0\rangle
\ee
The above relation, in turn, allows us to relate the electronic wave function $\mathcal{U}_e$ to the composite particle wave function $\mathcal{U}_\Phi$. 
\be
\mathcal{U}_e (\{z_i\}, \{\bar z_i\}) = \langle 0|\Psi_e(z_1, \bar z_1)...\Psi_e(z_N, \bar z_N)|N\rangle = \prod _{i<j}\left[\frac{\bar z_i - \bar z_j}{z_i - z_j}\right]^{m/2} \mathcal{U}_\Phi (\{z_i\}, \{\bar z_i\})\label{etophi}
\ee
Recalling that bosons can condense, $\mathcal{U}_\Phi = 1$ is an exact ground state of the mean field composite boson hamiltonian. In the case of the composite fermions, one recalls that (\ref{cflevel}) implies that the composite fermions have the correct degeneracy to fill exactly $p$ LLs. Thus $\mathcal{U}_\Phi = \mathcal{U}_{\Phi}^{p,B^*}$ is the exact ground state fo the composite fermion hamiltonian at the level of MFT. The resultant mean field electronic wave functions are:
\be
\mathcal{U}_e (\{z_i\}, \{\bar z_i\} = \langle 0|\Psi_e(z_1)...\Psi_e(z_N)|N\rangle = \prod _{i<j}\left[\frac{\bar z_i - \bar z_j}{z_i - z_j}\right]^{(2s+1)/2} \label{mfl}
\ee
when $p=1$ and $m$ is odd. And 
\be
\mathcal{U}_e (\{z_i\}, \{\bar z_i\}) = \prod _{i<j}\left[\frac{\bar z_i - \bar z_j}{z_i - z_j}\right]^{s} \mathcal{U}_{\Phi}^{p,B^*}(\{z_i\}, \{\bar z_i\})\label{mfj}
\ee
for the non-odd-integer fractions corresponding to $p \neq 1$. As noted in the introduction, the MF wave functions miss almost all crucial attributes of the FQH wave functions. The inadequacy of the MF results can be traced back to the fact that (\ref{fluxa}) needed to be a gauge transformation in order to respect the internal symmetries of the overall set up. This requirement  inevitably restricted the ``parameters" of (\ref{fluxa}) to be angles and the same angles ultimately reappeared as phases of the Jastrow factor in the MF wave functions.\\
One can recover the complete planar FQH wave functions by going beyond MFT e.g. following the hamiltonian approach pioneered in\cite{shankar}. Decomposing the density 
\be
\rho = \bar \rho + \tilde \rho \label{split}
\ee
into a MF part, $\bar \rho $, and a fluctuating component, $\tilde \rho$, \cite{fradkin, shankar} showed how $\tilde \rho$ contributions can be partially summed to recover (\ref{laughlin},\ref{jain}) from the mean field expressions obtained above\footnote{A brief summary of Shankar and Murthy's approach (adapated to complex space notation of this paper) is included in the Appendix for the sake of completeness.}. As argued before, this somewhat indirect way to turn fluxes to vortices, while certainly effective in the case of planar geometry, does not provide us with a framework for attaching vortices that might generalize to other geometries and problems. Further, the absence of  a natural parameter in the split (\ref{split}) renders the partial summation of $\tilde \rho$ effects somewhat uncontrolled. As hinted earlier, it is not unreasonable to expect that a better reorganization of the fundamental degrees of freedom might obviate the need to go beyond MFT. Applying the KKN framework to (\ref{He}) provides us with precisely such a reorganization allowing us to further trade the unitary gauge symmetry for a novel holomorphic symmetry (and fluxes for vortices), yielding a direct route to  (\ref{laughlin},\ref{jain}) within MFT itself. That is what we discuss next.
\section{Gauge Invariant Fields and Holomorphic Invariance}
We start with the second quantized electronic hamiltonian
\be
H_e = -\frac{2}{\mu}\int \Psi_e^\dagger (\bar \del + \bar A)(\del + A)\Psi_e. \label{he}
\ee 
$A = -\frac{i}{2}(A_1+iA_2)$ and $\bar A = -\frac{i}{2}(A_1-iA_2)$ are anti-hermitian complex combinations of the spatial components of the vector potential $A_i$. $A_0 = 0$ as is appropriate in a hamiltonian approach. The KKN framework utilizes the fact that spatial components of the gauge potentials can be parameterized in terms of a complex field $M$ as \cite{KKN1}
\be
A = -\partial M M^{-1} \hspace{.5cm} \bar A = M^{\dagger -1}\bar\del M^\dagger \label{M}
\ee
This parameterization holds even in the case of nonabelian gauge fields (in which case $M$ is also matrix in color space). As an operator, $M$ is a holomorphic Wilson line as is evident from its defining equation above which can be rewritten as $DM = 0$. $D$ is the usual gauge covariant derivative. The advantage of this parameterization is that general time-independent $local$ gauge transformations are realized as
\be
M \rightarrow U M
\ee
where $U$ is the unitary matrix implementing local gauge transformations. This implies that the field
\be
H = M^\dagger M
\ee
is gauge invariant. In the abelian case, one can parameterize $M$ in terms of a complex function $\phi$ as $M = e^\phi$, in which case (\ref{M}) implies the familiar gradient-curl decomposition of abelian gauge fields. The real and imaginary parts of $\phi$ yield the transverse and longitudinal parts respectively of the real gauge potentials $A_i$. We will continue to adhere to the complex notation as it is key to what we do later.\\
The parameterization (\ref{M}) has its own internal symmetry. The two fields $M$ and $M\bar V(\bar z)$ - where  $V(\bar z)$ is an arbitrary function depending only on the holomorphic coordinate $\bar z$ (and not on $z$) - yield the same gauge potential. In other words
\be
M(z,\bar z) \rightarrow M(z,\bar z)\bar V(\bar z)\label{hol}
\ee 
is a new $holomorphic$ symmetry that emerges from (\ref{M})\cite{KKN1}.\\
We can now use (\ref{M}) to recast (\ref{he}) in a gauge invariant form. There are two choices for generating  gauge invariant fermionic fields using $\Psi_e$ and $M$.
\be
{\mbox{Choice-I}}\hspace{1cm}\Lambda(z,\bar z) = M^{-1}(z,\bar z) \Psi_e(z,\bar z)
\ee
\be
{\mbox{Choice-II}}\hspace{1cm}\Lambda' (z,\bar z) = M^{\dagger}(z,\bar z) \Psi_e(z,\bar z)
\ee
Both $\Lambda $ and $\Lambda '$ are gauge invariant. The gauge invariant fields are composite operators formed by attaching holomorphic and anti-holomorphic Wilson lines to fermionic fields operators\footnote{Somewhat similar composite operators were also employed recently in the study of non-relativistic conformal field theories in \cite{tong}.}. Though gauge invariant, $\Lambda $ and $\Lambda '$  have nontrivial transformation properties under (\ref{hol}).
\be
\Lambda(z,\bar z) \rightarrow \bar V^{-1}(\bar z) \Lambda(z,\bar z), \hspace{.5cm} \Lambda(z,\bar z) ' \rightarrow V(z)\Lambda'(z,\bar z) \label{hol-l}
\ee
The gauge invariant canonically conjugate fields are:
\be
\tilde \Lambda  = \Psi^\dagger_e M = \Lambda ^\dagger H,\hspace{.5cm} \{\Lambda, \tilde \Lambda \} = 1 \label{mom}
\ee
\be
\tilde \Lambda'  = \Psi^\dagger_e M^{\dagger -1} = \Lambda ^{'\dagger} H^{-1},\hspace{.5cm} \{\Lambda', \tilde \Lambda' \} = 1
\ee
For later convenience we draw attention to the fact that the canonically conjugate fields are $not$ the same as the hermitian conjugates. The conjugate fields also transform nontrivially under holomorphic transformations.
\be
\tilde \Lambda(z,\bar z) \rightarrow \tilde \Lambda(z,\bar z)\bar V(\bar z), \hspace{.5cm} \tilde \Lambda '(z,\bar z) \rightarrow \tilde \Lambda'(z,\bar z)V^{-1}(z) 
\ee
The hamiltonian (\ref{he}) assumes the following form in terms of gauge invariant variables. Under \\{\bf Choice-I}:
\be
H_e \rightarrow H_\Lambda=  -\frac{2}{\mu}\int \tilde \Lambda (\bar \del + \bar J)\del \Lambda \label{hgi}
\ee
Every field appearing in the above hamiltonian is gauge invariant. The gauge fields have been replaced by a single complex current 
\be
\bar J = H^{-1}\bar \del H \label{j}
\ee
Under holomorphic transformations 
\be
\bar J \rightarrow \bar V^{-1} \bar J\bar V + \bar V^{-1}\bar \del \bar V
\ee
and it is easy to verify that the hamiltonian is invariant under (\ref{hol}). We have thus traded the unitary gauge symmetry for a new holomorphic symmetry in recasting the hamiltonian in terms of gauge invariant fields. Further, (\ref{hgi}) is self-adjoint, despite appearances. This may be verified using the nontrivial relationship between hermitian and canonical conjugates of $\Lambda$(\ref{mom}).\\
Since the adjoint of $\Lambda $ is $not$ its canonical conjugate, there is a nontrivial measure on the space of many-body wave functions. Given a general $N$ particle state 
\be
|N\rangle  = \frac{1}{\sqrt{N!}} \int \tilde{\mathcal{U}}_\Lambda(\{z_i\}, \{\bar z_i\})\tilde \Lambda(z_1, \bar z_1) \cdots \tilde\Lambda(z_N, \bar z_N)|0\rangle
\ee
$\langle N|N\rangle = 1$ implies that the wave functions are normalized as
\be
\int \prod_i H(z_i, \bar z_i) \tilde{\mathcal{U}}_\Lambda ^* \tilde{\mathcal{U}}_\Lambda = 1 \label{measure}
\ee
The overlaps of the wave functions must also be computed with the nontrivial measure; $\prod_i H(z_i, \bar z_i) $. The electronic wave functions are related to $\mathcal{U}_\Lambda$ as
\be
\mathcal{U}_e = \prod_i M(z_i, \bar z_i)  \tilde{\mathcal{U}}_\Lambda \label{etolambda}
\ee
which gives us back the flat measure on the space of electronic wave functions (as expected) using $H = M^\dagger M$. \\
A welcome consequence of the gauge invariant formalism is that the wave functions $\tilde{\mathcal{U}}_\Lambda$ are naturally in a Bargmann form\cite{girvin}. As we will see later, the factors of $M$ and $H$ in (\ref{etolambda}) and (\ref{measure}) will provide the gaussian factors needed for the FQH states and their overlaps respectively. The functions $\tilde{\mathcal{U}}_\Lambda$ themselves will be devoid of any gaussian factors.\\
Under \\{\bf Choice-II}:
\be
H_e \rightarrow H_{\Lambda '}=  -\frac{2}{\mu}\int \tilde \Lambda' (\del + J)\bar \del \Lambda' 
\ee
The complex current is now given by
\be
J = -\del H H^{-1} 
\ee
and it transforms as
\be
J \rightarrow V JV^{-1} - \del VV^{-1}
\ee
under holomorphic transformations. Once again the hamiltonian is self-adjoint but the norm on the wave functions is now set by $H^{-1}$.
\be
\int \prod_i H^{-1}(z_i, \bar z_i) \tilde{\mathcal{U}}_{\Lambda'} ^* \tilde{\mathcal{U}}_{\Lambda '} = 1 
\ee
The analog of (\ref{etolambda}) for the second choice is
\be
\mathcal{U}_e = \prod_i M^{\dagger -1}(z_i, \bar z_i)  \tilde{\mathcal{U}}_{\Lambda'}
\ee
There is a priori no reason to prefer one choice of gauge invariant variables over the other. However, within the context of any specific problem, physical consistency picks out a preferred choice. For example, in the context of relativistic field theories, the choice of gauge invariant variables can be resolved by appealing to compatibility with supersymmetry or anomaly conditions\cite{AKN}. In the present context, we shall see that requiring normalizability of wave functions will pick a unique choice of gauge invariant variables.\\ Before putting the gauge invariant framework to use, we note in passing that the currents $J$ (and $\bar J$), that play the role of gauge fields in the gauge invariant formalism are familiar entities in 2D conformal field theories\cite{KKN1}. If one defined a Wess-Zumino-Witten functional  with $H$ as the dynamical degree of freedom then $J$ (and $\bar J$) would be the associated left (and right) currents.
\section{Fractional Quantum Hall states on the Plane}
In this section we apply the holomorphic symmetry of the gauge invariant formalism to derive planar FQH states within a mean-field framework.
Starting with the hamiltonian (\ref{he}) we  restrict to the case of constant background magnetic field $F_{12} = B$, $B$ being a constant. The gauge choice $A_i = -\epsilon_{ij}\frac{B}{2}x_j$ ($\epsilon_{12} = -\epsilon_{21} = 1$)translates to $\bar A = -\frac{B}{4}z$, $A = \frac{B}{4}\bar z$. The matrices $M$ and $H$ can now be expressed as
\be
M(z,\bar z) = e^{-\frac{1}{4} Bz\bar z}, \hspace{.3cm} H(z,\bar z) = e^{-\frac{1}{2} Bz\bar z} \label{mh}
\ee
For later use, we also note that $H$ can be equivalently written as
\be
H = e^{-\frac{1}{2} Bz\bar z} = e^{-\frac{B}{2\pi}\int_w \ln [(z-w)(\bar z - \bar w)]}
\ee
The kernel in the exponential 
\be
G(z,\bar z;w,\bar w) = \ln[(z-w)(\bar z - \bar w)]
\ee
is the inverse laplacian in two spatial dimensions satisfying \footnote{It is understood that $\del _z \frac{1}{\bar z- \bar w} = \bar \del _z \frac{1}{z-w} =  \pi \delta ^2(z-w) $. $\delta ^2(z-w) \equiv \delta (Re(z) - Re(w))\delta (Im(z) - Im(w))$ }
\be
\del_z\bar\del_z G(z,\bar z;w,\bar w) =  \pi \delta ^2(z-w)  
\ee
We use {\bf Choice-I} to express (\ref{he}) in a gauge invariant form. The two choices correspond to two choices of the parity frame. Once the sign of $F_{12}$ is fixed, only one parity choice results in normalizable wave functions. As we shall justify later, {\bf Choice-I} is the one appropriate for our chosen sign of $B$. Under this choice (\ref{hgi})
\be
H_\Lambda =  -\frac{2}{\mu}\int \tilde \Lambda (\bar \del + \bar J)\del \Lambda \label{Hlambda}
\ee
and the holomorphic gauge potential
\be
\bar J = H^{-1}\bar\del H = -\frac{1}{2} B z, \hspace{.3cm} \del \bar J = - \frac{1}{2} B.
\ee
With the form of the gauge invariant hamiltonian now established, we can formulate holomorphic versions of composite boson and fermion theories to derive the Laughlin and Jain states.
\subsection{Composite Bosons}
It is intuitive to anticipate that the holomorphic version of (\ref{fluxa}) must result in a (holomorphic) statistical gauge potential that can be combined only with $\bar J$ (unlike (\ref{Hp}) the hermitian conjugate of $\bar J$ is not present in (\ref{Hlambda})). We accomplish this by mapping $\Lambda$ to a composite boson field $\Phi$ via the operator $\mathcal{R}$ as follows.
\be
\Lambda(z,\bar z) = e^{m\int_w \ln(\bar z - \bar w)\rho(w,\bar w)}\Phi(z,\bar z) = \mathcal{R}(z,\bar z)\Phi(z,\bar z)\label{readp}
\ee
The canonical conjugate field to $\Phi$ is $\Pi$, defined as
\be
\tilde \Lambda  = \Pi \mathcal{R}^{-1}
\ee
We take $m$ to be an odd integer to ensure $\Phi$ be bosonic. The canonical commutation relation between the composite fields is
\be
[\Phi(z), \Pi(z')] = \mathcal{R}^{-1}(z)\{\Lambda(z), \tilde\Lambda(z')\}\mathcal{R}(z') = \delta^2(z-z')
\ee
In contradistinction to (\ref{fluxa}), $\mathcal{R}$ defines a singular $holomorphic$ transformation (\ref{hol-l}) which is consistent with the emergent holomorphic symmetry of the gauge invariant formalism. Unlike the singular gauge transformation (\ref{fluxa}) the ``parameters'' of the field dependent holomorphic transformation do not have to be phases, as is evident in (\ref{readp}). The upshot is that  (\ref{readp}) attaches vortices instead of fluxes as will soon be clear. In fact $\mathcal{R}$ is the second quantized version of the Read\cite{read} operator \footnote{More precisely, $\mathcal{R}$ represents the nonlocal part of the Read operator without the exponential piece. The gaussian factor of the Read operator will be recovered in our framework separately.}. Proceeding further, we see that the density is mapped to itself under this map;  $\rho = \tilde \Lambda\Lambda = \Pi\Phi$. Further, the state $|0\rangle$ annihilated by $\Lambda$ is also annihilated by $\Phi$. 
The transformed hamiltonian can be expressed as
\be
H_\Lambda \rightarrow H_\Phi = -\frac{2}{\mu}\int \Pi(\bar \del + \bar J + \bar J_a)\del \Phi \label{cbh}
\ee
where the statistical holomorphic field
\be
\bar J_a: \del \bar J_a = m\pi \rho. \hspace{.5cm} \bar J_a(z) = \int_w \frac{m\rho (w,\bar w)}{\bar z - \bar w}  
\ee 
Once again, the hamiltonian is self-adjoint, which can be established using
\be
\Pi(z,\bar z) = \Phi^\dagger(z,\bar z) \mathcal{R}(z,\bar z)\mathcal{R}^\dagger(z,\bar z)H(z,\bar z) = \Phi^\dagger(z,\bar z)  e^{m\int_w \ln(z-w)(\bar z - \bar w )\rho(w,\bar w)}e^{-\frac{1}{2} Bz\bar z}
\ee 
We now consider a general state with $N$ composite bosons.
\be
|N\rangle = \frac{1}{\sqrt{N!}}\int_{z_1...z_N}\tilde{\mathcal{U}}_\Phi (\{z_i\}, \{\bar z_i\})\Pi(z_1, \bar z_1)...\Pi(z_N, \bar z_N)|0\rangle
\ee
The gauge invariant wave function corresponding to this state can be extracted by recasting the above equation as 
\begin{eqnarray}
|N\rangle &=& \frac{1}{\sqrt{N!}}\int_{z_1...z_N}\tilde{\mathcal{U}}_\Phi (\{z_i\}, \{\bar z_i\})\tilde \Lambda(z_1, \bar z_1)\mathcal{R}(z_1, \bar z_1)...\tilde \Lambda(z_N,\bar z_N)\mathcal{R}(z_N,\bar z_N)|0\rangle\nonumber\\
&= & \frac{1}{\sqrt{N!}}\int_{z_1...z_N}\tilde{\mathcal{U}}_\Phi (\{z_i\}, \{\bar z_i\})\prod _{i<j}(\bar z_i - \bar z_j)^m\tilde \Lambda(z_1,\bar z_1)...\tilde\Lambda(z_N,\bar z_N)\mathcal{R}(z_1,\bar z_1)...\mathcal{R}(z_N,\bar z_N)|0\rangle\nonumber\\
&=& \frac{1}{\sqrt{N!}}\int_{z_1...z_N}\tilde{\mathcal{U}}_\Phi (\{z_i\}, \{\bar z_i\})\prod _{i<j}(\bar z_i - \bar z_j)^m\tilde \Lambda(z_1,\bar z_1)...\tilde\Lambda(z_N,\bar z_N)|0\rangle \label{state-translation}
\end{eqnarray}
Thus, thought of as a gauge invariant electronic state, the wave function is
\be
\tilde{\mathcal{U}}_\Lambda (\{z_i\}, \{\bar z_i\}) = \langle 0|\Lambda(z_1,\bar z_1)...\Lambda(z_N,\bar z_N)|N\rangle = \prod _{i<j}(\bar z_i - \bar z_j)^m \tilde{\mathcal{U}}_\Phi (\{z_i\}, \{\bar z_i\})\label{bosel}
\ee
The overlap of this wave function is to be computed with respect to the measure given in (\ref{measure}). We can now go one step further and recover the gauge covariant electronic wave function using (\ref{etolambda}).
\beq
\mathcal{U}_e(\{z_i\}, \{\bar z_i\}) = \prod_i M(z_i, \bar z_i)  \tilde{\mathcal{U}}_\Lambda =  \prod _{i<j}(\bar z_i - \bar z_j)^m e^{-\frac{B}{4}\sum_i z_i\bar z_i }\tilde{\mathcal{U}}_\Phi (\{z_i\}, \{\bar z_i\}) \label{bosemain}
\eeq
No approximations have been made so far. We have recovered the full Jastrow factor as well as the correct gaussian as an exact factor that converts the composite fermion wave function to its electronic counterpart.\\ The above relation also makes it clear that the singular holomorphic transformation ($\mathcal{R}$) used to couple  composite bosons to the statistical gauge field attaches vortices instead of fluxes. We also note that the factors of $M$ transforming $\mathcal{U}_\Lambda$ to $\mathcal{U}_e$ provide the gaussian part of the Read operator, while $\mathcal{R}$ provides the Jastrow factor. \\
{\bf Holomorphic Mean Field Theory:} At this point we formulate a holomorphic MFT, i.e a mean field approximation tailored to the holomorphic description of the composite particles. We focus on the special bosonic state given by
\be
\mathcal{U}_\Phi = 1
\ee
which corresponds to a bose condensate where all the composite bosons are in the zero momentum state. In this state, the average density
\be
\bar\rho = \frac{N}{V} 
\ee
Further we tune the magnetic field such that
\be
m\bar\rho = \frac{B}{2\pi}\label{bosemf}
\ee
Invoking a mean-field approximation, we can replace $\rho$ by its expectation value $(\bar \rho)$, whence the composite boson hamiltonian (\ref{cbh}) simplifies to
\be
H_\Phi = -\frac{2}{\mu}\int \Phi^\dagger \bar\del\del \Phi 
\ee
The state corresponding to $\mathcal{U}_\Phi = 1$ is an exact zero energy eigenstate of the mean-field hamiltonian where the statistical gauge field experiences zero magnetic field. i.e. $\bar J + \bar J_a = 0$ since 
\be
\bar J_a(z) = \int_w \frac{m\rho (w,\bar w)}{\bar z - \bar w}   = \pi m \bar\rho z = - \bar J(z).
\ee
The electronic state corresponding to the bose-condensate  yields the Laughlin wave function $exactly$.
\be
\mathcal{U}_e(\{z_i\}, \{\bar z_i\}) =  \prod _{i<j}(\bar z_i - \bar z_j)^m e^{-\frac{B}{4}\sum_i z_i\bar z_i }\label{bosel}
\ee
{\bf Choice-I} vs {\bf Choice-II}: At this point it is useful to elaborate on picking between the two available options for gauge invariant variables. Use of the second choice (while keeping the sign of $F_{12}$ unaltered), would have resulted in poles instead of zeroes in the above wave function $(\bar z_i - \bar z_j) \rightarrow \frac{1}{(z_i - z_j)}$. It would have produced the wrong sign in the gaussian factor as well: $e^{-\frac{B}{4}\sum_i z_i\bar z_i } \rightarrow e^{\frac{B}{4}\sum_i z_i\bar z_i }$. The resulting wave function would have been non-normalizable and unphysical. As mentioned earlier, fixing the sign of $B$ amounts to a choice of the parity frame. Once that choice is fixed, there is a unique choice of gauge invariant variables that results in physically admissible results. Had we changed the sign of $B$ in (\ref{he}) at the outset, then {\bf Choice-II} would have been the physically meaningful option to select. The resulting wave function would have been the complex conjugate of (\ref{bosel}). Switching between the two choices amounts to $B \rightarrow -B$.
\subsection{Composite Fermions and Planar Jain States}
The composite fermion construction proceeds along similar lines as the bosonic case except $m (= 2s)$ must now be chosen to be an even number so that $\Phi$ represents a fermionic field.
\be
\{\Phi(z), \Pi(z') \} = \mathcal{R}^{-1}(z)\{\Lambda (z), \tilde \Lambda (z')\}\mathcal{R}(z') = \delta^2(z-z')
\ee
The hamiltonian remains (\ref{cbh}). The Jain states are obtained by tuning the constant magnetic field such that
\be
2\pi \bar \rho = B\nu, \hspace{.2cm} \nu = \frac{p}{2ps +1}\label{jain-fraction}
\ee
The map (\ref{bosemain}) also remains formally the same, though $\tilde {\mathcal{U}}_\Phi$ is now different. Further, since 
\be
\del \bar J + \del \bar J_a = -\frac{B}{2}\left(\frac{1}{2ps +1}\right),
\ee
the composite fermions experience an average magnetic field given by
\be
B^* = \frac{B}{2ps+1}
\ee
whence the degeneracy
\be
\frac{\bar \rho 2\pi}{B^*} = p
\ee
is just right to fill exactly $p$ Landau levels. Replacing $\rho$ by $\bar \rho$ in (\ref{cbh}) , we obtain the holomorphic mean field composite fermion hamiltonian.
\be
H_{\Phi} = -\frac{2}{\mu}\int \Pi(\bar \del - \frac{B^*}{2} z)\del \Phi, \hspace{.2cm} \Pi = \Phi^\dagger e^{-B^* z\bar z/2}\label{hcf}.
\ee
The Bargmann factor present in the relation between $\Phi^\dagger $ and $\Pi$ ensures that mean-field fermionic wave functions $ \tilde{\mathcal{U}}^1_\Phi$ and $ \tilde{\mathcal{U}}^2_\Phi$ are in the Bargmann form and devoid of gaussian factors. The gaussian factors needed for the computation of overlaps of composite fermion wave functions are incorporated in the Bargmann measure: $e^{-B^* z\bar z/2}$:
\be
(\tilde{\mathcal{U}}^1_\Phi, \tilde{\mathcal{U}}^2_\Phi) = \int e^{-B^* \sum z_i\bar z_i/2} \tilde{\mathcal{U}}_\Phi^{*1}(\{z_i\}, \{\bar z_i\}) \tilde{\mathcal{U}}^2_\Phi(\{z_i\}, \{\bar z_i\})
\ee
It is with respect this measure that (\ref{hcf}) is self-adjoint.\\
It is illustrative to look at the special case of $p=1$. Clearly, any holomorphic function of $\bar z_i$ can be regarded as a state in the LLL. In particular, the $N$ particle state 
\be
|N\rangle_{p=1}  = \frac{1}{\sqrt{N!}}\int_{z_i, \bar z_i} \Pi(z_1,\bar z_1)e^{\int_{w_1,\bar w_1} \ln(\bar z_1 - \bar w_1)\rho(w_1, \bar w_1)} \cdots \Pi(z_N,\bar z_N)e^{\int_{w_N,\bar w_N} \ln(\bar z_N - \bar w_N)\rho(w_N,\bar w_N)}|0\rangle\label{slater}
\ee
corresponds to the Slater determinant.
\be
\tilde{\mathcal{U}}_{\Phi}^{p=1, B^*} = \langle0|\Phi(z_N,\bar z_N)\cdots \Phi(z_1, \bar z_1)|N\rangle_{p=1} = \prod_{i<j}(\bar z_i - \bar z_j).
\ee
This is the antisymmetrized holomorphic ground state of (\ref{hcf}) when $p=1$. Using this in (\ref{bosemain}), yields the Laughlin wave function ($\nu = \frac{1}{2s+1}$) via the composite fermion picture within holomorphic MFT.
\beq
\mathcal{U}_e(\{z_i\}, \{\bar z_i\}) =\prod _{i<j}(\bar z_i - \bar z_j)^{2s+1} e^{-\frac{B}{4}\sum_i z_i\bar z_i } \label{p1}
\eeq
Higher composite fermion LLs may also be similarly constructed. To give an explicit example; the following unnormalized state belongs in the $k$-th LL:
\begin{eqnarray}
&|N\rangle_{k}  =\nonumber\\ &\int_{z_i, \bar z_i} e^{\int _{w,\bar w} \ln (w^k)\rho(w,\bar w)}\Pi(z_1,\bar z_1)e^{\int_{w_1,\bar w_1} \ln(\bar z_1 - \bar w_1)\rho(w_1,\bar w_1)} \cdots \Pi(z_N,\bar z_N)e^{\int_{w_N, \bar w_N} \ln(\bar z_N - \bar w_N)\rho(w_N,\bar w_N)}|0\rangle\nonumber
\end{eqnarray}
The corresponding wave function 
\be
\tilde{\mathcal{U}}_{{\Phi}}^k = \prod_{i< j}(\bar z_i - \bar z_j) \prod_l (z_l)^k 
\ee
is not holomorphic as it also involves $z_i$. Lack of holomorphicity is  a general feature of wave functions corresponding to $p > 1$. Explicit eigenstates of (\ref{hcf}) describing $p$ filled LLs can be obtained by removing the gaussian factors from known integer QH wave functions with $p$ filled LLs\footnote{We refer to \cite{jain-book} for several recipes for constructing wave functions corresponding to higher filled LLs. The explicit forms of these functions are not central to the present discussion.}.  Let us denote such eigenfunctions of (\ref{hcf}) by $\tilde{\mathcal{U}}_{\Phi}^{p, B^*} (\{\bar z_i\}, \{z_i\})$. Using (\ref{bosel}), we can now formally express gauge invariant wave electronic wave functions corresponding to $p>1$ as:
\be
\tilde{\mathcal{U}}_\Lambda^p = \prod _{i<j}(\bar z_i - \bar z_j)^{2s}\tilde{\mathcal{U}}_{\Phi}^{p,B^*} (\{\bar z_i\}, \{z_i\}).
\ee
Being devoid of gaussian factors, this wave function is in the Bargmann form as well.  As before, the factor $\prod_i H(z_i, \bar z_i) $ defines the Bargmann measure (\ref{measure}) on the state of gauge invariant wave functions.\\
Since $\nu <1$ we expect the gauge invariant state to lie in the LLL and the above expression clearly does not.  As is well known, this may be remedied by projecting the wave function to the LLL. It is gratifying to note that the projection also arises naturally in the gauge invariant framework. The presence of the Bargmann factor (\ref{measure}) in the measure implies that (for the purposes of computing overlaps of $\tilde{\mathcal{U}}_\Lambda^p$ with functions lying in the  LLL \cite{girvin}) the above wave function is equivalent to  
\be
\tilde{\mathcal{U}}_\Lambda^p = P_{LLL}\left[\prod _{i<j}(\bar z_i - \bar z_j)^{2s}\tilde{\mathcal{U}}_{\Phi}^{p,B^*} (\{\bar z_i\}, \{z_i\})\right]
\ee
where $P_{LLL}$ ``projects to the LLL" by replacing $z_i \rightarrow \frac{2}{B} \bar\del_i$. The implied ordering is that the derivatives are placed to the left of the expression. This projected state lies entirely within the LLL. We can now use (\ref{bosemain}) to finally obtain the $U(1)$ covariant electronic wave function
\be
\mathcal{U}_e^p(\{\bar z_i\}, \{z_i\}) =    e^{-\frac{B}{4}\sum_i z_i\bar z_i } P_{LLL}\left[\prod _{i<j}(\bar z_i - \bar z_j)^{2s}\tilde{\mathcal{U}}_{\Phi}^{p,B^*} (\{\bar z_i\}, \{z_i\})\right] \label{holwf}
\ee
which corresponds to Jain states.\\ At this point it is worth emphasizing a few  takeaway points.\\ 
{$\bullet$}The gauge invariant analysis along with the underlying holomorphic symmetry has allowed us to attach vortices to particles and construct the second quantized Read operator leading to a derivation of both the Laughlin and Jain states as exact results within holomorphic mean field theory. \\
{$\bullet$} The $P_{LLL}$ operation also follows naturally from the Bargmann form of the wave functions obtained within the gauge invariant framework and no additional arguments are needed for the derivatives (in the $P_{LLL}$ operation) to not act on the gaussian part of the wave function.\\
{$\bullet$} In the Appendix we present an argument showing that the Bohm-Pines method applied to the gauge invariant set-up does not yield any further fluctuation driven corrections to (\ref{holwf}). 
\section{Fractional Quantum Hall states on $S^2$}
In this section we extend the previous analysis to $S^2$. Issues related to holomorphicity are best analyzed by regarding $S^2$ as $CP^1 = SU(2)/U(1)$. The coset space representation of the sphere has been used extensively in the study of higher dimensional integer quantum hall effects\cite{KN1, KN2} and gauge fields on $R\times S^2$ \cite{AN2}. We shall refer to these papers for a fuller exposition of relevant technical details. For the sake for completeness, we gather some essential geometric aspects necessary for the present paper. General $SU(2)$ group elements can be parameterized as
\be
g = \left(\begin{array}{cc}
\bar u_2 & u_1\\
-\bar u_1 & u_2
\end{array}\right)  = \frac{1}{\sqrt{1+z\bar z}} \left(\begin{array}{cc}
1& z\\
-\bar z & 1
\end{array}\right) \left(\begin{array}{cc}
e^{i\psi/2} & 0\\
0& e^{-i\psi/2}
\end{array}\right) \label{g}
\ee
Global ``spinor" $SU(2)$ coordinates satisfy $|u_1|^2 + |u_2|^2 = 1$. $z,\bar z$ provide local stereographic (dimensionless) coordinates  on $S^2$ and they are related to $u_\alpha$ by $z = u_1/u_2$. $\psi$ parameterizes the $U(1)$ which is quotiented out of $SU(2)$ to produce the round sphere. Setting $\psi = 0 $ in the defining equations for $u_\alpha$ yields the spinorial coordinates used in Haldane's seminal paper \cite{haldane}.  The metric on $CP^1$ is the Fubini-Study metric.
\be
g_{z\bar z} = g_{\bar z z} =  \frac{4r^2}{(1+z\bar z)^2}
\ee
 such that the volume element is\footnote{Once again it is understood that the dimensionless complex coordinate $z$ decomposes as $z = x-iy$ and $d^2z$ is a shorthand for the flat measure $dxdy$.}
\be
d\mu = \frac{4r^2 d^2z}{(1+z\bar z)^2}.
\ee
The total volume of the sphere being $V = 4\pi r^2.$\footnote{We use $\mu$ to denote both electronic mass and the spherical volume element. We hope the difference will be clear from the context.}\\
A basis for functions on $SU(2)$ is provided by the Wigner $\mathcal{D}^j_{pq}$ functions: $\mathcal{D}^j_{pq} = \langle j,p|\hat g|j,q\rangle$. The left and right translation operators on the Lie group are defined by
\be
L_a g = \frac{1}{2}\sigma _a g, \hspace{.3cm} R_a g = g\frac{1}{2}\sigma _a \label{l-r}
\ee
where $\sigma_a$ are the Lie algebra generators. The complex combinations $L_\pm = L_i \pm i L_2$ and $R_\pm = R_i \pm i R_2$ obey  $SU(2)$ commutation relations.
\be
[R_+, R_-] = 2R_3, \hspace{.3cm} [L_+, L_-] = 2L_3. 
\ee 
In the present context $L_\pm = L_i \pm i L_2$ and $R_\pm = R_i \pm i R_2$ will play the role of spherical analogs of magnetic and spatial translation operators respectively. It is also understood that $p$ and $q$ in $\mathcal{D}^j_{pq}$ represent $L_3$ and $R_3$ eigenvalues respectively.\\ Scalar functions on $S^2$ correspond to null $R_3$ charge; i.e $\mathcal{D}^j_{p0}$, which are proportional to the usual spherical harmonics. \\
Single particle observables can be defined using expressions for $L_a$ and $R_a$ as differential operators, which can be obtained  using (\ref{l-r}). In particular
\be
R_+ = \bar u_2\partial _{u_1} - \bar u_1\partial_{u_2}, \hspace{.3cm} R_- = u_2\partial _{\bar u_1} -  u_1\partial_{\bar u_2}\label{rg}
\ee
When acting on scalar functions on $S^2$, these reduce to the familiar form
\be
R_+ = (1+ z\bar z)\partial_z, \hspace{.3cm} R_- = (1+z\bar z)\partial_{\bar z}\label{rl}
\ee
At the risk of digressing we point out the following important subtlety. The expressions (\ref{rg}) are valid on any $SU(2)$ valued function. However the expressions for the right translation operators in local coordinates (\ref{rl}) are only valid when $R_\pm$ act on scalar functions. When acting on vectorial spherical quantities which carry  nonzero $R_3$ charges, (\ref{rl}) must be replaced by appropriately covariantized versions, $\mathcal{R}_\pm$. As an example, if we take a spherical function $f(z,\bar z)$, a short calculation shows:
\be
R_-f =  (u_2\partial _{\bar u_1} -  u_1\partial_{\bar u_2})f = \frac{u_2}{\bar u_2}(1+z\bar z)\partial_{\bar z}f 
\ee
The factor $\frac{u_2}{\bar u_2}$ shows that the action of the lowering operator takes us out of the coset $SU(2)/U(1)$ and the resulting function is no longer a scalar defined on a sphere. The covariantized form of $R_+$ in local coordinates can be computed using the expression for $R_+$ in global coordinates.
\be
R_+\frac{u_2}{\bar u_2} \tilde f(z,\bar z) = \bar u_2\partial _{u_1} - \bar u_1\partial_{u_2}(\frac{u_2}{\bar u_2} \tilde f(z,\bar z)) = ((1+ z\bar z)\partial_z - \bar z)\tilde f(z,\bar z)
\ee
The last term yields the covariantized expression for the raising operator in local coordinates.
\be
R_+ \rightarrow \mathcal{R}_+ = (1+ z\bar z)\partial_z - \bar z
\ee
Similarly
\be
R_- \rightarrow \mathcal{R}_- = (1+ z\bar z)\partial_{\bar z} - z
\ee
The extra terms in the covariantized expressions above are nothing but components of the spin connection on $S^2$.\\
With these geometric details in place, we can express the spherical electronic hamiltonian as:
\be
H_e = -\frac{1}{2\mu}\int d\mu \Psi^*_e \left( \frac{1}{r}\mathcal{R}_{-} + A_{-}\right)\left( \frac{1}{r}R_{+} + A_{+}\right)\Psi_e \label{hes}
\ee
The frame fields $A_\pm$ are spherical versions of the complex planar gauge potentials $A, \bar A$. As on the plane, we can parameterize spherical gauge fields in terms of a complex matrix $M$ as
\be
A_+ = -\frac{1}{r}(R_+M)M^{-1}, \hspace{.3cm} A_- = M^{\dagger -1}\frac{1}{r}R_-M^\dagger
\ee
It is now clear that 
\be
M \rightarrow M\bar V(\bar u_1, \bar u_2)
\ee
is a symmetry of the parameterization. This is the spherical counterpart of the planar holomorphic symmetry encountered earlier.\\ As in the planar case, we transform to the gauge invariant field ($\Lambda$) corresponding to {\bf Choice-I}\footnote{Since the sign of the magnetic field will be kept the same as the planar analysis, we do not present details corresponding to {\bf Choice-II} on $S^2$.}
\be
\Lambda = M^{-1} \Psi_e,
\ee
leading to the gauge invariant hamiltonian
\be
H_e \rightarrow H_\Lambda = -\frac{1}{2\mu}\int d\mu \tilde \Lambda \left(\frac{1}{r}R_-  + J_-\right)\frac{1}{r}R_+\Lambda.\label{HSL}
\ee
\be
J_- = H^{-1}\frac{1}{r}R_-H; \hspace{.3cm} H = M^\dagger M
\ee
Once again, all fields appearing in the hamiltonian above are gauge invariant, though they transform nontrivially under holomorphic transformations. 
\be
J_- \rightarrow \bar V^{-1} J_- \bar V + \bar V^{-1} R_-\bar V, \Lambda \rightarrow \bar V^{-1}\Lambda, \tilde \Lambda \rightarrow \tilde \Lambda \bar V \label{hol-s}
\ee
We can now specialize to the case of constant background magnetic field for which the frame fields are given by monopole gauge potentials.  
\be
A_+ = \frac{n\bar z}{2r}, \hspace{.3cm} A_- = - \frac{nz}{2r}\label{MA}
\ee
The dimensionless number $n$ is an integer and the gauge fields satisfy the Dirac quantization rule
\be
\frac{1}{2\pi}\int d\mu F_{12} = n.
\ee
$n$ is related to the real magnetic field $B$ by
\be
n = 2r^2B.
\ee
Further, in the monopole background:
\be
M = e^{\frac{n}{2}\ln{u_2\bar u_2}} = e^{-\frac{n}{2}\ln (1+z\bar z)}, \hspace{.3cm} H = e^{n\ln{u_2\bar u_2}}. \label{ms}
\ee
 The holomorphic gauge potential assumes the form
 \be
J_- = -\frac{nz}{r}.
\ee
As a consistency check we note that setting $z = \frac{v}{2r}$ and taking the large $r$ limit (\ref{hes}) reduces to its familiar planar counterpart 
\be
H_e \rightarrow -\frac{2}{\mu}\int d^2v \Psi^*_e(\partial_v + B\frac{\bar v}{4})(\partial_{\bar v} - B\frac{v}{4})\Psi_e
\ee
where $v,\bar v$ depict the planar complex coordinates. It is easy to see that the planar limits of $M, H$ and (\ref{HSL}) are also correctly recovered in the same limit.
\subsection{Vortex Attachment on $S^2$}
So far everything has closely paralleled the planar construction. However, the construction of the Read operator requires us to tackle new subtleties that are unique to the sphere. The planar inverse laplacian - $\ln[(z-w)(\bar z - \bar w)]$ - split into holomorphic and antiholomorphic parts, and the holomorphic part, $\ln(\bar z - \bar w)$, was used to construct the Read operator (\ref{readp}). However, the inverse spherical laplacian in local $S^2$ coordinates is
\be
G = \ln \left[ \frac{(z-w)(\bar z - \bar w)}{(1+z\bar z)(1+ w\bar w)}\right].
\ee
Clearly this expression does not decompose into holomorphic and antiholomorphic components in local stereographic coordinates. Additionally, the equation satisfied by $G$:
\be
\mathcal R_+(z) R_- (z)G(z,w) = \mathcal R_-(z)R_+(z)G(z,w) = 4\pi r^2\delta ^2_\mu(z-w) - 1\label{zerog}
\ee
differs from the expression on the plane in an important way. The first term on $rhs$ parallels the equation obeyed on the plane.\footnote{$\delta ^2_\mu(z-w)$ denotes the delta function normalized to the measure on the sphere. $\delta ^2_\mu(z-w)  = \frac{1}{4r^2}\delta^2(z-w)(1+z\bar z)^2$.} However, the second term has to do with the subtraction of the zero mode of the laplacian. This additional contribution does not have a planar analog.\\ 
Let us address the factorization issue first. Denoting the global coordinates corresponding to $z$ and $w$ by $u_\alpha$, $u'_\alpha$ respectively, we note that $G$ can equivalently be expressed as
\be
G =  \ln \left[ \frac{(z-w)(\bar z - \bar w)}{(1+z\bar z)(1+ w\bar w)}\right] = \ln (\epsilon_{\alpha \beta} u_\alpha u'_\beta) + \ln (\epsilon_{\alpha \beta} \bar u_\alpha \bar u'_\beta).
\ee
Thus, the inverse laplacian decomposes into its holomorphic and antiholomorphic components in global coordinates. This decomposition suggests that the  composite particle field $\phi$ be defined via a singular holomorphic transformation as follows.
\be
\Lambda  = \left[\exp m\int d\mu(u') \ln(r\epsilon_{\alpha \beta} \bar u_\alpha \bar u'_\beta)\rho(u',\bar u')\right] \phi = \mathcal{R} \phi \label{reades}
\ee
The integration above is over the stereographic spherical coordinates and $\psi$ for $u'$ (\ref{g}) has been set to zero. As far as $u$ is concerned, the corresponding $\psi$ coordinate can take on any fixed value and without  loss of generality, it can be set to zero as well.
The resulting hamiltonian in local coordinates is
\be
H_\Lambda \rightarrow H_\phi = -\frac{1}{2\mu}\int \Pi_\phi \left[\frac{1}{r} \mathcal R_- - \frac{nz}{r} + \frac{m}{r} \int d\mu(w) \frac{1+z\bar w}{\bar z - \bar w}\rho(w,\bar w)\right]\frac{1}{r}R_+\phi \label{hs1}.
\ee
However, $H_\phi$ does $not$ provide a satisfactory description of composite particles attached to vortices. Its shortcoming may be seen by analyzing the 
putative statistical holomorphic gauge field 
\be
\bar j_a(z,\bar z) =  \frac{m}{r} \int d\mu(w) \frac{1+z\bar w}{\bar z - \bar w}\rho(w,\bar w)\label{jrho},
\ee
which satisfies (using (\ref{zerog}))
\be
\frac{1}{r} \mathcal{R}_+\bar j_a(z,\bar z) = 4\pi m \rho(z,\bar z) - \frac{m}{r^2}\int d \mu \rho
\ee
The last term on the $rhs$ is due to the zero mode of the scalar laplacian and it brings us to the second subtlety mentioned previously. This term was absent on the plane and its presence on the sphere poses some obvious impediments. It prevents us from tuning the average density to cancel all or a part of $J_-$ at the mean field level. It also stands in the way of interpreting the above equation as a constitutive relation implied by a statistical  Chern-Simons coupling. This issue can be remedied by recalling that the zero mode contribution of the spherical scalar laplacian can be removed by inserting additional monopole fields \cite{ouvry}. To this end we define a second holomorphic transformation that redefines the composite particle field as
\be
\phi(i_a) = \left[\exp -(m(N-1)\ln \bar u_2^a)\right] \Phi(i_a) \label{mp2}.
\ee
The exponentials insert monopole operators (of strength $m(N-1)$) at the positions of composite particles described by the field $\Phi$, which is a scalar field on $S^2$.\\  
We can now construct $N$ particle states as\footnote{We denote the $CP^1$ coordinates $(u_\alpha^a, \bar u_\alpha ^a)$ collectively by $i_a$.} 
\be
|N\rangle = \frac{1}{\sqrt{N!}}\int \tilde{\mathcal{U}}_\Phi(i_1\cdots i_N)e^{-m(N-1)\ln \bar u_2^1}\Pi_\phi(i_1)\cdots e^{-m(N-1)\ln \bar u_2^N}\Pi_\phi(i_N)|0\rangle. \label{sp-state}
\ee
Equivalently:
\be
|N\rangle = \frac{1}{\sqrt{N!}}\int \tilde{\mathcal{U}}_\Phi(i_1\cdots i_N)\Pi_\Phi(i_1)\cdots \Pi_\Phi(i_N)|0\rangle. \label{sp-state-2}.
\ee
$\Pi_\Phi$ is the momentum operator conjugate to $\Phi$. It is understood that the $CP^1$ coordinates are integrated over.\\
One can also understand the insertion of the additional monopole fields ($\phi \rightarrow \Phi$) on purely geometric grounds. For $\Lambda $ to be a function on $S^2$, it must be annihilated by $R_3$\footnote{$R_3 = \frac{1}{2}(u_\alpha \partial _{u_\alpha} - \bar{u}_\alpha \partial _{\bar{u}_\alpha}) $}. However, $\mathcal{R}$ as defined in (\ref{reades}) is an exact eigenstate of $R_3$ with a  nonzero eigenvalue.
\be
R_3\mathcal{R} = \left(-\frac{m}{2}\int \rho \right) \mathcal{R}\label{r3}
\ee
Thus, for (\ref{reades}) to be consistent, $\phi$ must transform nontrivially under $R_3$ to compensate for (\ref{r3}). (\ref{mp2}) ensures just that. With $\Phi$ taken to be a function on $S^2$, it is easy to see that (\ref{mp2}) guarantees that $R_3\Lambda = 0$ at the mean field level.\\
Finally, the composite particle hamiltonian can now be expressed in terms of these new variables ($H_\phi \rightarrow H_\Phi$):
\be
H_\Phi = -\frac{1}{2\mu}\int \Pi_\Phi \left[\frac{1}{r} \mathcal R_- - \frac{nz}{r} + \frac{m}{r} \int d\mu(w) \frac{1+z\bar w}{\bar z - \bar w}\rho(w,\bar w) + \frac{m}{r}z(N-1)\right]\frac{1}{r}R_+\Phi \label{HCBS2}
\ee
This is the hamiltonian that we will use to describe bosonic and fermionic composite particles on $S^2$ and analyze at the mean field level in the next subsection.\\ In preparation for MFT, it is useful to note that  
the last term inside the parenthesis in $H_\Phi$ is the additional monopole field generated by (\ref{mp2}) and it precisely cancels the zero mode contribution of the spherical Green's function in MFT. The resultant statistical gauge field (including the additional monopole term) is given by:
\be
\bar J_a(z,\bar z) =  \frac{m}{r} \int d\mu(w) \frac{1+z\bar w}{\bar z - \bar w}\rho(w,\bar w) + \frac{m}{r}z(N-1).\label{sp-j}
\ee
Further, it is instructive to note that that monopole term $\frac{m}{r}z(N-1) = \frac{m}{r}z\int d\mu \bar \rho $. Specifically, for a system of $N$ particles, the mean-field density
\be
\bar \rho = \frac{N-1}{4\pi r^2}\label{meanrho}.
\ee
The subtraction of $1$ from $N$ is the result of subtracting a divergent self interaction contribution. This is best understood in the first quantized language, where $\rho$ enters  the first quantized version of ($\ref{hs1}$) via the interaction term $\sim \sum_i \int_w \frac{1+z_i\bar w}{\bar z_i - \bar w}\rho(w,\bar w)$. Subtraction of the singular self interaction terms implies that $\rho(w,\bar w) \sim \sum_{j\neq i} \delta ^2(w-z_j)$ $\Rightarrow$ $\int \rho \sim N-1$. The difference between $N$ and $N-1$ is not particularly significant on the plane where the Hilbert space is effectively infinite dimensional, but it is crucial on the sphere where the Hilbert space is truly finite dimensional.\\
 (\ref{sp-state-2},\ref{HCBS2} and \ref{sp-j}) are all the ingredients we need to describe the dynamics of the composite particles on $S^2$. Next we analyze holomorphic mean field theories for the bosonic and fermionic cases respectively.
\subsection{Composite Bosons} Setting $m=2s+1$ once again implies that the composite particles are bosons. If one now tunes the magnetic field such that \footnote{The three expressions are equivalent to one another.}
\be
4\pi m \bar \rho = \frac{n}{r^2}, \hspace{.2cm} \mbox{or}\hspace{.2cm} m\bar \rho = \frac{B}{2\pi}, \hspace{.2cm}\mbox{or}\hspace{.2cm} m(N-1) = n.\label{bcb}
\ee
the composite bosons experience zero magnetic field on average since
\be
\mathcal{R}_+(\bar J_a(z,\bar z) + J_-(z,\bar z)) = 0\label{zerob}.
\ee
 This can be verified by replacing $\rho $ by its mean field value - $\bar \rho $ - in (\ref{sp-j}). We can now replicate the steps following (\ref{state-translation}) and translate  composite particle wave functions to electronic wave functions. Commuting all the factors of $\mathcal R$ to the right and using the fact that $\mathcal R \equiv 1$ when acting on the vacuum, we can express (\ref{sp-state}) as
\be
|N\rangle = \frac{1}{\sqrt{N!}}\int \tilde{\mathcal{U}}_\Phi (i_1\cdots i_N)\prod_{a<b} \left( r\epsilon_{\alpha \beta}\bar u^{a}_\alpha \bar u^{b}_\beta\right)^m\prod_c (\bar u ^c_2)^{-m(N-1)}\tilde \Lambda (i_1)\cdots \tilde \Lambda (i_N)|0\rangle
\ee
Thus
\be
\tilde{\mathcal{U}}_\Lambda (i_1\cdots i_N) = \langle 0|\Lambda(i_1)\cdots \Lambda (i_N)|N\rangle  = \prod_{a<b} \left( r\epsilon_{\alpha \beta}\bar u^{a}_\alpha \bar u^{b}_\beta\right)^m\prod_c (\bar u ^c_2)^{-m(N-1)}  \tilde{\mathcal{U}}_\Phi (i_1\cdots i_N)
\ee
We can pull out a factor of $\bar u^a_2\bar u^b_2$ from every term in the antisymmetrized product which exactly cancels the monopole term $\prod_c (\bar u ^c_2)^{-m(N-1)} $. As a result, we can we can express the wave function manifestly in terms of the local $S^2$ coordinates as follows.
\be
\tilde{\mathcal{U}}_\Lambda (\{z_i\}, \{\bar z_i\})  = \prod_{i<j} \left(r(\bar z_i - \bar z_j)\right)^m  \tilde{\mathcal{U}}_\Phi (\{z_i\}, \{\bar z_i\})\label{lambda-s}
\ee
Finally, following (\ref{bosemain}) and using (\ref{ms}) for the spherical version of the matrix $M$, we can relate the electronic and composite particle wave functions explicitly. 
\be
\mathcal{U}_e (\{z_i\}, \{\bar z_i\})  = \prod_a (1+\bar z_a z_a)^{-\frac{n}{2}}\prod_{i<j} \left(r(\bar z_i - \bar z_j)\right)^m  \tilde{\mathcal{U}}_\Phi (\{z_i\}, \{\bar z_i\})\label{bose-s}
\ee
Setting $\tilde{\mathcal{U}}_\Phi = 1$ in (\ref{bose-s}) yields the spherical analog of the odd integer Laughlin wave function. As on the plane, this state corresponds to a condensate of composite bosons and it is an exact zero-energy mean field ground state of (\ref{HCBS2}). It is also gratifying to note that setting $z = \frac{v}{2r}$ in (\ref{bose-s}) and taking $r\rightarrow \infty$, we recover the planar expression (\ref{bosel}); up to overall unimportant numerical factors.
\subsection{Composite Fermions and Jain states}
Spherical Jain states may be obtained by first rendering the composite particles fermionic; $m = 2s$. Following the planar analysis, we require that the composite fermions experience an effective magnetic field $B^*$ that is just right for $p$ filled LLs. We indicate the corresponding effective monopole strength by $n^*$, i.e.
\be
n^* = 2B^*r^2
\ee
Further, we recall that the degeneracy of the $p$th LL is $n+2p+1$. Thus, if
\be
n^* = \frac{N}{p} -p,\label{effn}
\ee
the composite fermions would have the right degeneracy to fill exactly $p$ LLs. If the original monopole strength is tuned such that
\be
n = \left(\frac{2ps +1}{p}\right)N - (2s +p)\label{effmon}
\ee
then using the mean  field density (\ref{meanrho}) in  (\ref{HCBS2}) we see that:
\be
\frac{1}{r}\mathcal{R}_+(\bar J_a + J_-) = -\frac{1}{r^2}n^*.
\ee
The above equation shows that (\ref{effmon}) generates the required $B^*$ corresponding to (\ref{effn}). The spherical mean field composite fermion hamiltonian (the spherical counterpart of (\ref{hcf})) is
\be
H_{\Phi} = -\frac{1}{2\mu}\int \Pi_\Phi \left[\frac{1}{r} \mathcal R_- - \frac{n^*z}{r} \right]\frac{1}{r}R_+\Phi. 
\ee
At this point, can follow the planar analysis to construct spherical versions of the Jain states. The odd integer (Laughlin) states correspond to $p=1$ where the composite fermions fill the LLL. Explicitly; the equivalent of (\ref{slater}) on the sphere is 
\begin{eqnarray}
&|N\rangle_{p=1}  = \\
&\int_{z_i, \bar z_i} \Pi_\Phi(z_1,\bar z_1)e^{\int_{w_1,\bar w_1} \ln(r(\bar z_1 - \bar w_1))\rho(w_1,\bar w_1)} \cdots \Pi_\Phi(z_N,\bar z_N)e^{\int_{w_N,\bar w_N} \ln(r(\bar z_N - \bar w_N))\rho(w_N,\bar w_N)}|0\rangle\label{slaters}\nonumber
\end{eqnarray}
The resultant composite fermion wave function is, once again, the Slater determinant.
\be
\tilde{\mathcal{U}}_{\Phi}^{p=1} = \langle0|\Phi(z_N,\bar z _N)\cdots \Phi(z_1,\bar z_1)|N\rangle_{p=1} = \prod_{i<j}r(\bar z_i - \bar z_j)
\ee
Transforming back to electronic variables by repeating the steps leading to (\ref{lambda-s},\ref{bose-s}), we recover the Laughlin-Haldane wave function for the odd integer LLs.
\be
\mathcal{U}_{e}^{p=1} =  \prod_a (1+\bar z_a z_a)^{-\frac{n}{2}}\prod_{i<j} \left(r(\bar z_i - \bar z_j)\right)^{2s+1}
\ee
For other values of $p$, the composite fermion wave functions $\tilde{\mathcal{U}}^{p,n^*} _\Phi$ correspond to higher LLs. For our purposes, what matters is not their explicit form but the fact that they are known\footnote{$\tilde{\mathcal{U}}^{p,n^*} _\Phi$ can be obtained by removing the geometric factors $(1+z\bar z)^{-n^*/2}$ from known spherical expressions for filled Landau levels.}. Formal expressions for electronic wave functions corresponding to $p>1$ can once again be obtained following (\ref{lambda-s},\ref{bose-s}).
\be
\mathcal{U}_{e}^{p} =  \prod_a (1+\bar z_a z_a)^{-\frac{n}{2}}P_{LLL}\left[\prod_{i<j} \left(r(\bar z_i - \bar z_j)\right)^{2s}  \tilde{\mathcal{U}}^{p, n^*}_\Phi(\{z_i\},\{\bar z_i\}) \right]\label{jains}
\ee
As in the planar case, the projection operation follows from the Bargmann form of the wave functions. For spherical geometry, the projection operation amounts to rewriting the argument of the projection operator in terms of the $CP^1$ coordinates and replacing $u_\alpha \rightarrow \frac{\partial}{\partial \bar u_\alpha}$\footnote{Details of the projection operation as well as the accompanying ordering issues are discussed in depth in \cite{jain-book}.}.\\
(\ref{jains}) is nothing but the spherical counterpart of the Jain wave function which the present analysis has produced as an exact result entirely within MFT. This result is meant to serve as a proof of principle that the vortex attachment technique discussed in this paper extends to nontrivial geometries as does the notion of the emergent holomorphic symmetry. Finally,  by writing $z = \frac{v}{2r}$ one can check that (\ref{jains}) goes smoothly over to its planar limit ($v$ being the complex dimensionful planar coordinate) in the $r\rightarrow \infty$ limit.
\section{Concluding Remarks} The central result of this paper is a derivation of planar and spherical Laughlin and Jain wave functions as exact results in (holomorphic) MFT. Key to this economical route to the celebrated wave functions is a reformulation of  nonrelativistic electronic dynamics using gauge invariant variables first used in\cite{KKN1}. This reformulation  trades the received gauge symmetry for an emergent holomorphic symmetry. The resultant complex framework yields a direct analytical method for attaching vortices to particles which makes it possible to construct the second quantized Read operator and derive both planar and spherical FQH wave functions entirely within holomorphic MFT. Additionally, the $P_{LLL}$ operation also follows as consequence of the fact that the gauge invariant wave functions are in the Bargmann which, in turn, follows from self-adjointness requirements within the gauge invariant hamiltonian formalism.\\
One may wonder if it is reasonable to expect the simplifications listed above to follow from what is fundamentally only a change of dynamical variables. An analogy may be made to the problem of computing the nonperturbative mass-gap in pure Yang-Mills theory in D=2+1. This theory is known to be massless to any finite order in perturbation theory. However a gauge invariant reformulation of the theory \cite{KKN1} immediately yields a gapped spectrum and provides an analytical lower bound on the mass gap. In effect, the gauge invariant reorganization of  degrees of freedom automatically incorporates effects of processes that amount to summing over an infinite number of Feynman diagrams. In fact, a self consistent but approximate resummation of Feynman diagrams contributing to the mass gap of the gauge theory \cite{AN} is known to compare well with the exact analytical hamiltonian result obtained in\cite{KKN1}. Returning to the present context, the analogous advantage of the gauge invariant reformulation is to automatically incorporate the effects of fluctuation dependent processes allowing us to bypass beyond-MFT analyses \cite{fradkin,shankar}) that are required to turn fluxes into vortices within the usual gauge covariant framework.\\
Several immediate future directions of inquiry follow from the methods presented in this paper.\\
 It would undoubtedly be interesting to apply the present methods to study excited FQH states and hole excitations and to the larger question of particle-hole symmetry in FQH systems \cite{son}. \\
In the relativistic regime, the Dirac equation with vortex gauge potentials and its connection to the relativistic version of the Laughlin wave function was recently investigated in \cite{dolan}. Extending the present work to relativistic fermions may yield an alternative approach to the study of such systems.\\
Prompted by the impressive body of work \cite{KN1, KN2} on the integer quantum Hall effect in general $CP^k$ spaces, it is tempting to ask if a higher dimensional generalization of the methods introduced in this paper may be possible.\\
Finally, the issue of particle-hole symmetry in FQH systems has led recently to much interesting work on particle-vortex dualities \cite{son, MV,CS,MA}. Further, the gauge invariant  techniques and variables used in this paper were recently shown to play an important role in realizing particle-vortex dualities within a hamiltonian framework \cite{AAL}. It is thus natural to ask what bearing the analytical technique for attaching vortices introduced here may have on the broader issue of particle-vortex dualities.\\
We hope that future work can shed light on some of the intriguing questions listed above.\\
{\bf Acknowledgements:} We are deeply grateful to Sriram Ganeshan, Pouyan Ghaemi, Dimitra Karabali, V.P.Nair  and Alexios Polychronakos for illuminating discussions and for their feedback on an earlier version of this draft.
\section{Appendix} In this appendix we summarize the hamiltonian method for improving upon the mean field results (\ref{mfl}, \ref{mfj}) pioneered by Shankar and Murthy. We adapt the analysis of \cite{shankar, shankar-book} to the complex coordinates used in this paper and show that the same method applied to the holomorphic mean field wave function does not yield any corrections to (\ref{holwf}).\\
If one applies the decomposition of the density into a mean field value, $\bar \rho$,  and a fluctuating  component $\tilde \rho$ (\ref{split}) to the constitutive relations (\ref{fluxpin}) and retains $\tilde \rho$, then then the resulting composite particle hamiltonian (\ref{compH}) takes on on the following form.
\be
H_\Phi = -\frac{2}{\mu}\int \Phi^\dagger(\bar \partial - \frac{z}{4}B^* + \frac{m}{2}\int _w \frac{\tilde \rho (w,\bar w)}{\bar z - \bar w})(\partial + \frac{\bar z}{4}B^* - \frac{m}{2}\int _w \frac{\tilde \rho (w,\bar w)}{z - w})\Phi\label{hfluc} 
\ee
The ``beyond MFT" contributions are encoded in the fluctuation ($\tilde \rho $) dependent terms\footnote{The use of complex coordinates also has the advantage of turning real space the``inverse curl" operator highlighted in \cite{shankar} into the more tame 2D Dirac propagators $(z-w)^{-1},  ({\bar z - \bar w})^{-1}$ which are very familiar e.g. in the context of 2D conformal field theories.}. We will restrict ourselves to the more general composite fermion ($m = 2s$) case in this discussion as that covers both the Laughlin and Jain states. Shankar and Murthy's approach to dealing with this hamiltonian involved enlarging the Hilbert pace by introducing a new real scalar field $\alpha $ and its conjugate momentum $\Pi_\alpha \equiv -i \frac{\delta}{\delta \alpha}$. Physical states are annihilated by $\alpha $
\be
|\Psi\rangle_{physical} : \alpha |\Psi\rangle_{physical}  = 0\label{phys}
\ee
The hamiltonian was further modified to
\be
H_\Phi = -\frac{2}{\mu}\int \Phi^\dagger(\bar \partial - \frac{z}{4}B^* +  \bar{\mathcal{A}}+ \frac{m}{2}\int _w \frac{\tilde \rho (w,\bar w)}{\bar z - \bar w})(\partial + \frac{\bar z}{4}B^* + \mathcal{A} - \frac{m}{2}\int _w \frac{\tilde \rho (w,\bar w)}{z - w})\Phi\label{haug} 
\ee
Where the vector fields 
\be
\mathcal{A} = - \partial \alpha, \hspace{.3cm} \bar{\mathcal{A}} = \bar \del \alpha. 
\ee
are transverse.
Clearly (\ref{haug}) is equivalent to (\ref{hfluc}) on states that satisfy (\ref{phys}). The new degrees of freedom are introduced to remove the fluctuation terms via a unitary transformation generated by
\be
U = \exp\left[i\frac{m}{2}\int_{z,w} \Pi_\alpha (z,\bar z) \ln \left[(z-w)(\bar z - \bar w)\right]\tilde \rho (w,\bar w)\right],
\ee
which acts on the fields as follows.
\begin{eqnarray}
U^\dagger \bar{\mathcal{A}}(z,\bar z)U &=& \bar{\mathcal{A}}(z,\bar z) - \frac{m}{2}\int _w \frac{\tilde \rho (w,\bar w)}{\bar z - \bar w}\nonumber \\
U^\dagger \mathcal{A}(z,\bar z)U &=& \mathcal{A}(z,\bar z) + \frac{m}{2}\int _w \frac{\tilde \rho (w,\bar w)}{ z -  w}\nonumber \\
U^\dagger \Phi (z,\bar z)U &=& \exp(i\frac{m}{2}\int_w \ln \left[(z-w)(\bar z - \bar w)\right]\Pi_\alpha (w)) \Phi (z,\bar z)
\end{eqnarray}
In deriving the last equation, it has been assumed that $\tilde \rho = :\Phi^\dagger \Phi:$. The unitary transformation removes the dependence of $H$ on $\tilde \rho$ at the expense of introducing $\Pi_\alpha$. The transformed hamiltonian $H' = U^\dagger H U$ is.
\be
H' = -\frac{2}{\mu}\int \Phi^\dagger(\bar \partial - \frac{z}{4}B^* +  \bar{\mathcal{A}}+ \frac{im\pi}{2}\int _w \frac{\Pi_\alpha(w,\bar w)}{\bar z - \bar w})(\partial + \frac{\bar z}{4}B^* + \mathcal{A} + \frac{im\pi}{2}\int _w \frac{\Pi_\alpha(w,\bar w)}{z - w})\Phi \label{hun}
\ee
The physicality condition is also transformed under the unitary transformation to:
\be
\left(\alpha (z) - \frac{m}{2}\int_w \ln (z-w)(\bar z - \bar w)\tilde\rho(w,\bar w))  \right)|\Psi\rangle_{Physical} = 0\label{physun}
\ee
Shankar and Murthy identified that the term in $H'$ that is quadratic in the auxiliary fields can account for corrections to the wave function. We denote it by $H'_0$.
\be
H'_0 = + \frac{2\bar \rho}{\mu}\int_z\left( \bar \del \alpha(z,\bar z) \del \alpha(z,\bar z)  + \frac{m^2 \pi^2}{4}\int _w \frac{\Pi_\alpha(w,\bar w)}{\bar z - \bar w}\int _w \frac{\Pi_\alpha(w,\bar w)}{z - w}\right)\label{osch}
\ee
In the above expression, another approximation has been made, namely the replacement of $\Phi^\dagger \Phi$ by the mean field density $\bar \rho$. The exact (zero energy) ground state of $H'_0$ can be easily obtained. It is given by.
\be
\Psi '_0 = \exp \left [ \frac{1}{m\pi} \int \alpha \del \bar \del \alpha \right] = \exp \left [ \int _{z,z'}\frac{m}{4}\tilde \rho(z,\bar z)\ln \left[(z-z')(\bar z - \bar z')\right]  \tilde \rho(z',\bar z')\right)\label{oscgs}
\ee
(\ref{physun}) has been used above to obtain the expression for the ground state in terms of densities. Finally, Shankar and Murthy express 
\be
\tilde \rho(z,\bar z)  = \rho(z,\bar z) - \bar \rho = \sum_i\delta^2(z-z_i) - \bar \rho
\ee
Using this in (\ref{oscgs}), the argument of the exponential splits into two terms $I_1$ and $I_2$ (apart from an unimportant constant contribution).
\be
I_1 = \frac{m}{4}\sum_{i,j}\int_{z,z'} \delta^2(z-z_i)\ln \left[(z-z')(\bar z - \bar z')\right] \delta^2(z' - z'_j) = 2s\sum_{i<j}\ln |\bar z_i - \bar z_j|
\ee 
\be
I_2 =-\frac{m\bar \rho}{2}\sum_{i}\int_z \ln \left[(z-z_i)(\bar z - \bar z_i)\right] = -s\bar \rho\pi \sum_i z_i\bar z_i = -\frac{B}{4}\frac{2ps}{2ps+1} \sum_i z_i\bar z_i 
\ee
We have used (\ref{jain-fraction}) to eliminate $\bar \rho$ in terms of $B$ in the last expression above. Thus
\be
\Psi '_0 = \prod_{i<j}|\bar z_i - \bar z_j|^{2s}\exp\left(-\frac{B}{4}\frac{2ps}{2ps+1} \sum_i z_i\bar z_i \right) \label{osc-final}
\ee
This correction multiplies the mean field result (\ref{mfj}). Thus the corrected electronic wave function is
\begin{eqnarray}
\mathcal{U}_e (\{z_i\},\{\bar z_i\}) &=& \prod _{i<j}\left[\frac{\bar z_i - \bar z_j}{z_i - z_j}\right]^{s} \mathcal{U}_{\Phi}^{p,B^*}(\{z_i\},\{\bar z_i\}) \Psi '_0(\{z_i\},\{\bar z_i\}) \nonumber\\
&=&  \prod _{i<j} (\bar z_i - \bar z_j)^{2s} \exp\left((-\frac{B}{4}\frac{2ps}{2ps+1} - \frac{B^*}{4})\sum_i z_i\bar z_i \right)\tilde{\mathcal{U}}_{\Phi}^{p, B^*}(\{z_i\},\{\bar z_i\})\nonumber \\
&=& \prod _{i<j} (\bar z_i - \bar z_j)^{2s} \exp\left(-\frac{B}{4}\sum_i z_i\bar z_i \right)\tilde{\mathcal{U}}_{\Phi}^{p,B^*}(\{z_i\},\{\bar z_i\})\nonumber 
\end{eqnarray}
In the second step above, the gaussian has been factored out of $\mathcal{U}_{\Phi}^{p,B^*}$, and the final result is the unprojected Jain wave function.\\
It is worth emphasizing again at this point that no details specific to the system hosting the FQH states were used in the derivation above. This strongly suggests that the Bohm-Pines procedure  resums generic physical effects which were missed at the MFT level due to the choice of the original degrees of freedom and the corresponding unitary internal symmetry. Reformulating the theory using holomorphic gauge invariant variables addresses this issue and it leads to the complete Jain and Laughlin wave functions as exact results entirely within MFT. \\
It is worth investigating what happens if the above approach were to be applied to the holomorphic hamiltonian (\ref{hcf}). To verify that the Bohm-Pines approach does not produce further corrections to (\ref{holwf}), one can can include the fluctuating density term in (\ref{hcf}) and express the complete hamiltonian as
\be
H_{cf} = -\frac{2}{\mu}\int \Pi(\bar \del - \frac{B^*}{2} z +m\int _w \frac{\tilde \rho (w)}{\bar z - \bar w})\del \Phi
\ee
Since the symmetry is now no longer unitary, only the holomoprhic field $\bar{\mathcal{A}}$ need be included in the hamiltonian to remove the density. The augmented hamiltonian is
\be
H_{cf} = -\frac{2}{\mu}\int \Pi(\bar \del - \frac{B^*}{2} z +\bar{\mathcal{A}} + m\int _w \frac{\tilde \rho (w)}{\bar z - \bar w})\del \Phi
\ee
The above hamiltonian is not self-adjoint in general, but it is so on physical states satisfying (\ref{phys}).
The transformation that removes density term is now no longer unitary. Instead we require a holomorphic similarity transformation generated by
\be
S = \exp\left[im \int_{z,w}\Pi_\alpha(z,\bar z)\ln(\bar z - \bar w) \tilde \rho (w,\bar w)\right].
\ee
The transformed hamiltonian $H' = S^{-1}H_{cf}S$ is
\be
H' =  \int \Pi(\bar \del - \frac{B^*}{2} z +\bar{\mathcal{A}} + im\int _w \frac{\Pi_\alpha(w,\bar w)}{\bar z - \bar w})\del \Phi\label{aughol}
\ee 
while the physicality condition is 
\be
\left(\alpha (z,\bar z) - m\int_w \ln(\bar z - \bar w)\tilde\rho(w,\bar w))  \right)|\Psi\rangle_{Physical} = 0
\ee
The term quadratic in the auxiliary fields (\ref{osch}) that generated  corrections to the mean field wave function is simply not present in the holomorphic case (\ref{aughol}). This shows that the Bohm-Pines method applied to the holomorphic framework does not produce further corrections to (\ref{holwf}). 

\end{document}